\documentclass[twocolumn,showpacs,preprintnumbers,amsmath,amssymb,aps,prb]{revtex4}
\usepackage{graphicx}
\begin{document}

\title{
Pinning, Ordering, and Dynamics of Vortices in Conformal Crystal 
and Gradient Pinning Arrays 
} 
\author{
D. Ray, C. Reichhardt, and C. J. Olson Reichhardt 
} 
\affiliation{
Theoretical Division,
Los Alamos National Laboratory, Los Alamos, New Mexico 87545 USA
} 

\date{\today}
\begin{abstract}
We numerically investigate magnetization, pinning, ordering, and
dynamics of vortices interacting with pinning arrangements which have
a density gradient. 
We focus on conformal crystal structures
obtained by conformally transforming a spatially uniform
periodic array, 
as well as non-conformal gradient structures and structures with 
quasiperiodic order.
The conformal structures feature a density gradient 
and local ordering. 
Using magnetization simulations we find that
conformal pinning arrays exhibit enhanced pinning 
compared to non-conformal gradient arrays as well as 
compared to random, periodic, and quasiperiodic arrays, 
for a broad range of fields.
The effectiveness of conformal arrays arises from
the continuum of length scales introduced into the arrays by the 
conformal transformation, 
allowing for a broad range of local commensuration effects.  
At higher vortex fillings above the range of conformal effectiveness, 
we show that a non-conformal rectangular gradient array exhibits
strong pinning 
due to a novel commensuration effect and vortex ordering. 
Using transport simulations where vortices are driven along 
the gradient and at an angle to the gradient, 
we confirm the effectiveness of conformal pinning at increasing the
critical current. 
For a rotated drive, the gradient
arrays produce a strong vortex guidance effect in the direction
perpendicular to the gradient.  
\end{abstract}
\maketitle

\vskip2pc

\section{Introduction}
One of the most important issues for many applications of type-II
superconductors is understanding how to optimize the pinning of
magnetic flux vortices, since the system becomes dissipative when
vortices are in motion \cite{R}.  The applied current at which
vortices depin is called the critical current
$J_c$, and there has been
intense research in tailoring the pinning in a superconductor 
so as to maximize $J_c$
for either specific applied fields or for a broad
range of fields.  
Beyond the direct application to superconductivity, 
vortex pinning and dynamics provide a very rich system for exploring
collective ordering of particles on random or periodic substrates, as
well as for studying nonequilibrium phenomena at depinning which arise
in a variety of condensed matter systems, such as driven charge
density waves \cite{G}, sliding Wigner crystals \cite{C}, atomic
friction \cite{L}, adhesion \cite{A}, magnetic domain wall motion
\cite{M}, and driven colloidal systems \cite{Colloids}.

A widely used method for enhancing the pinning and critical
current in superconductors is the creation of nanostructured
artificial arrays of pinning sites \cite{1,2,3,4,5,6}, featuring a
well-defined number of sites arranged in a desired configuration.
These sites pin vortices effectively, but only a limited number of sites
can be added to a sample
since their creation causes
local damage to the superconductor that decreases the
material's current-carrying capacity.  Therefore, it is of great
importance to determine the optimal geometrical arrangement
of a {\it fixed} number of pinning sites
that generates the strongest
pinning over the widest range of fields.  In random site
arrangements, where there is a distribution of interpin spacings,
the closely-spaced pins are inefficiently used due to strong
short-range intervortex repulsion, while widely-spaced pins 
permit easy flow of vortices between them \cite{channel}.  Periodic
pinning arrays such as triangular or square arrays, which avoid both 
these shortcomings, have been studied in detail; the pinning is enhanced
compared to random arrays, but only near certain commensurate field
values, where the number of vortices is an integer
\cite{1,2,3,4,5,6,7,8,9} or fractional \cite{10,11} multiple of the
number of pinning sites.  At commensurability, ordered or partially
ordered vortex arrangements form which can have various types of
symmetries \cite{2,7,8,9,10,11}; however, away from commensurability,
the vortex structures are partially or totally disordered,
allowing for the formation of weak spots and easy vortex flow channels
which reduce the depinning threshold \cite{2,7,10}.  Rectangular
pinning arrays that are periodic with two different length scales 
have been shown to produce anisotropic vortex structures and transport
\cite{12,13,14,15,16}.

There have been a number of efforts
to enhance the pinning in periodic arrays under
noncommensurate conditions.  Systematically diluting a triangular
pinning array gives enhanced pinning not only at matching fields, but
also at non-integer matching fields corresponding to what the integer
matching fields for the non-diluted array would have been
\cite{17,18,19,20,21,22}.
Other studies showed that the addition of some disorder to a
triangular pinning lattice enhances
the pinning at incommensurate fillings, 
but at the cost of reducing the maximum pinning at
the integer matching fields \cite{23}.  
Asymmetric system geometries, such
as funnel geometries \cite{24,25,26} or periodic constriction
geometries \cite{27} which utilize vortex jamming,  
produce new types of commensurability effects. 
Vortex artificial ice states can be created using
superlattices of
square pinning arrangements \cite{28,29,30}, 
and exhibit strong matching at certain
non-integer fields.  Composite pinning lattices based on 
intermeshed periodic arrays have also been created \cite{31,32,33}.

Another approach to pinning enhancement is the use of pinning
structures that combine aspects of periodicity and disorder.  Studies
of quasiperiodic arrays where pinning sites were placed at the
vertices of a Penrose tiling \cite{34,35,36,37,38,39,40} revealed
novel non-integer matching conditions, and showed that the
overall pinning can be enhanced over periodic triangular pinning
arrays below the first matching field
\cite{34,35,37,39}.  

Despite these efforts, 
pinning arrays which rely on commensuration effects 
have continued to possess the fundamental drawback 
of only providing enhanced pinning near 
certain field values, rather than over a broad range of field. 
In an effort to overcome this shortcoming, 
a new type of pinning geometry, known as a conformal crystal
pinning array, was recently proposed\cite{41};  
an example of such an array is shown in Fig.~\ref{fig:transf}(b).
A conformal crystal is constructed by performing a conformal, or
angle-preserving, transformation upon a uniform lattice structure such
as a hexagonal array, producing a new structure which preserves
the {\it local} ordering of the original array but now
exhibits a density gradient in one direction \cite{42,41}.  Conformal
crystals were first observed as the approximate ground state structure
of a set of repulsively interacting magnetic spheres restricted to
move in two dimensions and subjected to a gravitational potential;
the distinctive arching appearance of the resulting
structure inspired the whimsical nickname of ``gravity's rainbow'' \cite{42}.  
In vortex simulations, conformal pinning arrays produced enhanced
pinning compared to an equivalent number of pinning sites arranged
randomly with uniform density over a wide range of applied fields, 
with the enhancement persisting until the applied field
was increased beyond the maximum local pinning density present in the
array \cite{41}. 
The conformal array also 
outperformed uniform periodic arrays except in the immediate vicinity of
the matching field. 
A random array with a density gradient
produced pinning only marginally better than a non-gradient random array 
and substantially worse than the conformal array, 
indicating that the preservation of local order 
was as vital to the conformal array's performance 
as was the introduction of a gradient.  
The enhancement was observed both in flux gradient-driven
simulations \cite{41}, where the vortices enter or exit the system from
the edges, and in current-driven simulations \cite{41,driven}, where the
vortex density is fixed and the vortices are
driven across the sample with a uniform Lorentz force.  
Two subsequent experiments confirmed the enhanced pinning in conformal
arrays.  In the first\cite{43},
conformal pinning arrays displayed enhanced critical currents 
over uniform periodic pinning arrays 
except in the vicinity of certain matching fields.
In the second\cite{44}, conformal pinning arrays 
showed decreased magnetoresistance in
transport experiments 
compared to periodic or diluted periodic structures 
with the same number of pinning sites.

Pinning arrays with a density gradient not produced by 
a conformal transformation have also been investigated. 
A pinning array consisting of concentric 
rectangular ``rings'' of pinning sites with a gradually increasing
separation between the rings was fabricated and shown to exhibit
enhanced pinning compared to a periodic pinning array \cite{45}.
Additionally, there have been numerical studies of pinning arrays arranged
in circular hyperbolic tessellation structures so that the pinning sites
have a gradient \cite{46}; however, these studies did not include
comparisons with periodic or random pinning arrays to check whether
the hyperbolic array produces enhanced pinning over the other
geometries.

In this work, we use molecular dynamics simulations to 
study magnetization, depinning, ordering and dynamics of 
vortices in systems with various types of pinning arrays featuring 
a density gradient. We investigate the effect of different local 
orderings in conformal crystal arrays by considering transformed 
versions of hexagonal, square, quasiperiodic, 
and Archimedean \cite{51,archi} lattices. 
We also consider two non-conformal systems, a
rectangular pinning array with a one-dimensional (1D) gradient, 
and a random pinning arrangement with a density gradient. 
This manuscript is organized as follows: 
in Section II, we describe how conformal arrays are created and 
enumerate the various pinning arrays which will be analyzed. 
In Section III, we explain our physics model and the types of 
simulations we perform. 
In Section IV, we show the results of magnetization simulations, 
demonstrating 
that conformal pinning arrays produce
enhanced pinning over non-conformal gradient arrays 
over a wide range of applied fields, 
as well as over arrays of uniform density such as the quasiperiodic 
Penrose array or periodic arrays (aside from a narrow range around 
matching fields).  
In Section V, 
by examining the pattern of occupied pinning sites 
as the field is increased, we show that conformal arrays are effective 
due to a {\it local} commensuration effect which is an extension  
of the global matching exhibited by uniform pinning arrays: 
because the conformal transformation introduces a continuously changing 
lattice constant into the array, 
the vortex lattice can 
progressively match with different slices of the array as the 
number of vortices is increased. 
Consequently, 
we find that the particular type of local order present is relatively
unimportant in determining the effectiveness of conformal pinning arrays,
as long as some type of ordering is present 
that would give a matching effect in a uniform array.
We also use local commensuration to explain the 
maximum effective field value for conformal arrays. 
At higher fields well above this value,  
we show 
in Section VI 
that a rectangular gradient pinning array 
gives enhanced pinning over the conformal
arrays due to a novel commensuration effect 
which gives rise to 
ordered vortex chain structures.  
In Section VII, 
we examine current-driven vortex dynamics
for vortices driven along the gradient direction 
as well as at varied angles with respect to the gradient.
We find that in general, conformal pinning arrays reduce
vortex motion compared to the other arrays, 
over the full range of drive angles.
Compared to the random and rectangular gradient arrays 
which exhibit channels of easy vortex flow, 
the arching structure of the conformal arrays
prevents the formation of such channels 
for any choice of drive angle. 
Additionally we find that many gradient arrays 
have a pronounced vortex guidance effect, 
where vortices move much more easily perpendicular to the gradient 
than along the gradient.
Finally, in Section VIII, we present our conclusions.

\section{Gradient Pinning Arrays}

\subsection{Conformal Arrays}
Conformal transformations are a well-studied topic from complex analysis
\cite{cbook}. 
An analytic function $w=f(z)$ maps two infinitesimal vectors 
originating from the same point in the $z$ plane with angle $\theta$ 
between them to two new vectors in the $w$ plane separated by the same 
angle. We use this property to transform an ordered structure into 
one which remains locally ordered but which is distorted on a global scale.

\begin{figure}
\includegraphics[width=\columnwidth]{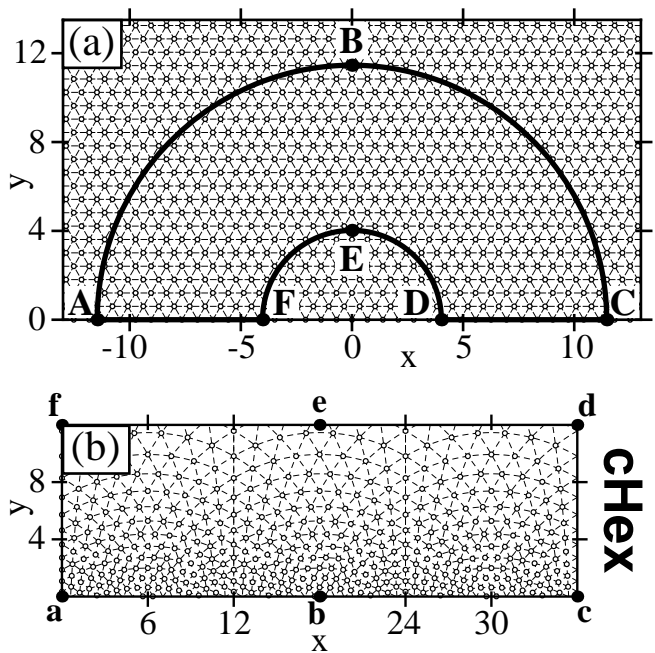}
\caption{ 
(a) A semiannular section of a uniform hexagonal lattice. 
(b) Effect of a conformal transformation upon the region in (a). 
The resulting structure (conformal hex or ``cHex'') has local 
hexagonal ordering and a density gradient. 
See text for details of the transformation. 
}
\label{fig:transf}
\end{figure}

In this work, we allow the transformation
\begin{equation*}
w=\frac{\pi}{2\alpha}+\frac{1}{i\alpha} \ln(i\alpha z)
\end{equation*}
(where $\alpha$ is a constant which we set to $\pi/36$) to act on a
semiannular section of the $z$-plane specified by $\mathrm{Im} z\geq
0$ and $r_{\mathrm{in}} \leq |z| \leq r_{\mathrm{out}}$ for some choice
of $r_{\mathrm{in}}$ and $r_{\mathrm{out}}$.  This transformation maps
radial lines in the $z$ plane to vertical lines in the $w$ plane, and
circular arcs centered at the origin to horizontal lines. 
Consequently, 
the semiannular region {\bf ABCDEF} 
is mapped to the rectangle {\bf abcdef}, as shown in
Fig.~\ref{fig:transf}. 
The circular arcs in the $z$ plane are increasingly
stretched by the transformation as their radial coordinate decreases:
fixing $\alpha r_{\mathrm{out}}=1$, arc {\bf ABC} has the same length as
line {\bf abc}, while arc {\bf DEF} is stretched to the same 
length as lines {\bf def} and {\bf abc}. 
This increasing stretch causes an ordered structure with
uniform density, such as the hexagonal lattice 
shown in Fig.~\ref{fig:transf}(a), 
to be mapped to a structure with a density gradient, 
as shown in Fig.~\ref{fig:transf}(b).  
The local
ordering of the original structure is preserved due to the
conformal nature of the mapping: each site in the transformed lattice
retains its six nearest neighbors.

\begin{figure}
\includegraphics[width=\columnwidth]{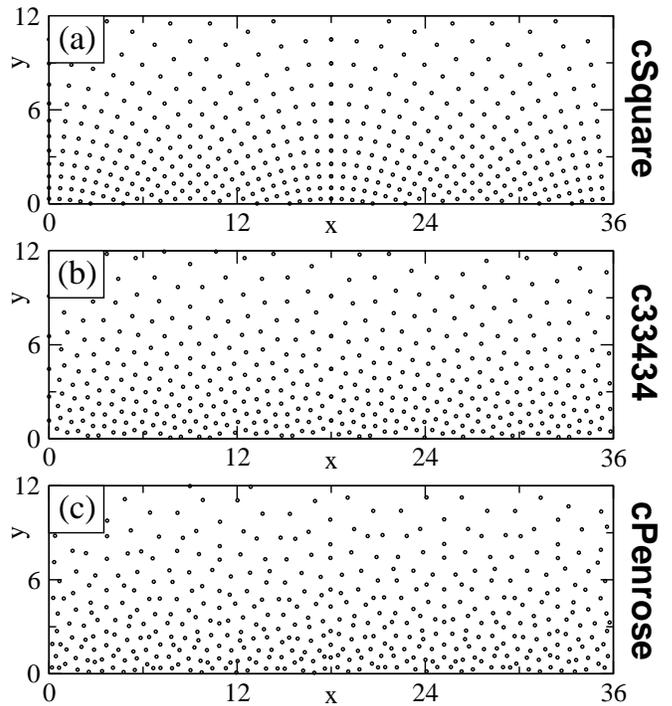}
\caption{ 
Conformally transformed versions of various uniform (non-gradient) 
structures: 
(a) square lattice; 
(b) Archimedean 33434 lattice (a periodic tiling of the plane using 
squares and triangles);
(c) 5-fold Penrose tiling using thin and thick rhombs.
}
\label{fig:carray}
\end{figure}

The conformally transformed hexagonal structure was studied
in previous work \cite{41,driven}, where it was termed a
``conformal pinning array'' or ``CPA.'' Since in the present work 
we also consider other types of local order, we here refer to this array 
as a ``conformal hex'' array, or ``cHex'' for short. 
Figure \ref{fig:carray} illustrates
conformal arrays with other types of local order.
In Fig.~\ref{fig:carray}(a), a transformed square lattice 
exhibits local 4-fold 
order. We refer to this tiling as ``conformal square'' or ``cSquare.'' 
Fig.~\ref{fig:carray}(b) shows the transformed version of a 
so-called ``Archimedean 
33434'' array. Archimedean tilings of the plane are periodic tilings created
using two or more types of regular polygon.
Several such tilings exist \cite{archi}; here we consider a particular 
type, 33434, composed of squares and triangles. 
A uniform pinning array based on such a tiling has been studied 
extensively \cite{51}.  We refer to the transformed version as 
``conformal 33434'' or ``c33434''. 
Finally, Fig.~\ref{fig:carray}(c) 
shows the transform of a Penrose tiling with approximate 
5-fold symmetry, constructed using thin and thick rhombs \cite{penrose}. 
Pinning arrays based upon such Penrose tilings have been simulated 
previously \cite{34}; here, we consider transformed Penrose arrays 
(``conformal Penrose'' or ``cPenrose''), as well as a uniform 
(untransformed) Penrose array.
The uniform version of each array type will be referred to, when 
required, as ``uHex'', ``uSquare'', ``u33434'', and ``uPenrose.''

All pinning arrays fit into a rectangular region of dimension 
$36\lambda \times 12\lambda$, where $\lambda$ is the London penetration
depth, and have
a nominal average density of pinning sites equal to 
$\bar{\rho}_p=1.0/\lambda^2$ 
(i.e. each array contains approximately 432 pins, with small 
deviations on the order of 1\% due to boundary effects.) 
This density can be obtained
through appropriate choices of the parameter $r_{\mathrm{in}}$ and 
the lattice constant of the original uniform lattice.
The local pinning density varies from a maximum of 
$\rho_p^{\rm loc}=2.0/\lambda^2$ 
to a minimum of $\rho_p^{\rm loc}=0.3/\lambda^2$. 

\subsection{Arrays produced by Other Methods}

\begin{figure}
\includegraphics[width=\columnwidth]{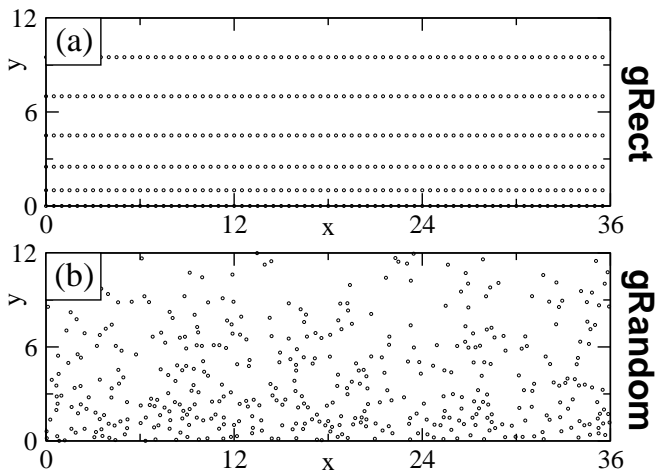}
\caption{ 
Gradient pinning arrays not formed by a conformal transformation.  
(a) A rectangular gradient pinning array, consisting of rectangles stretched 
and compressed in one dimension. 
(b) Random (disordered) pinning with a density gradient matching 
the conformal arrays.
}
\label{fig:garray}
\end{figure}

The conformal arrays described above all 
have the common features of local ordering and a
density gradient. The local pinning density (or equivalently, the
local lattice constant) varies fairly smoothly through the
array due to the continuous nature of the conformal
transformation. In order to determine the relative importance of each
of these two features 
in maximizing the effectiveness of the pinning, we consider two other
array types.

The first alternative array is shown in Fig.~\ref{fig:garray}(a) 
and consists of a rectangular
pinning array with a lattice constant that varies in the gradient
direction so as to approximate the density profile of the conformal
arrays as closely as possible. We call this array ``rectangular with
gradient,'' or ``gRect.''  The gRect array differs from the
conformal arrays in two respects: ($i$) the lattice constant does not
vary smoothly through the array, and ($ii$) the original uniform
rectangular array from which the gradient array was constructed does
not exhibit a commensuration peak. The importance of this second
property will be explained in Section \ref{sec:localcomm}.

The second alternative array, shown in Fig.~\ref{fig:garray}(b), 
has a density gradient matching 
that of the conformal arrays, but the pinning sites are arranged 
randomly subject only to this constraint, so that all local ordering 
is eliminated. This array, called ``random with gradient'' or 
``gRandom,'' was studied in earlier work \cite{41} and found 
to be only minimally more effective than uniform random pinning. 
Thus, we use the gRandom array to establish a baseline of minimum 
effectiveness for pinning arrays with a density gradient. 
Notice that this array type can be generated either by using the 
desired density profile as a bias when generating pinning locations, 
or by conformally transforming a uniform random array; the 
transformation does not add order where none was originally present.

\section{Simulation and System}   
\begin{figure*}
\includegraphics[height=3in]{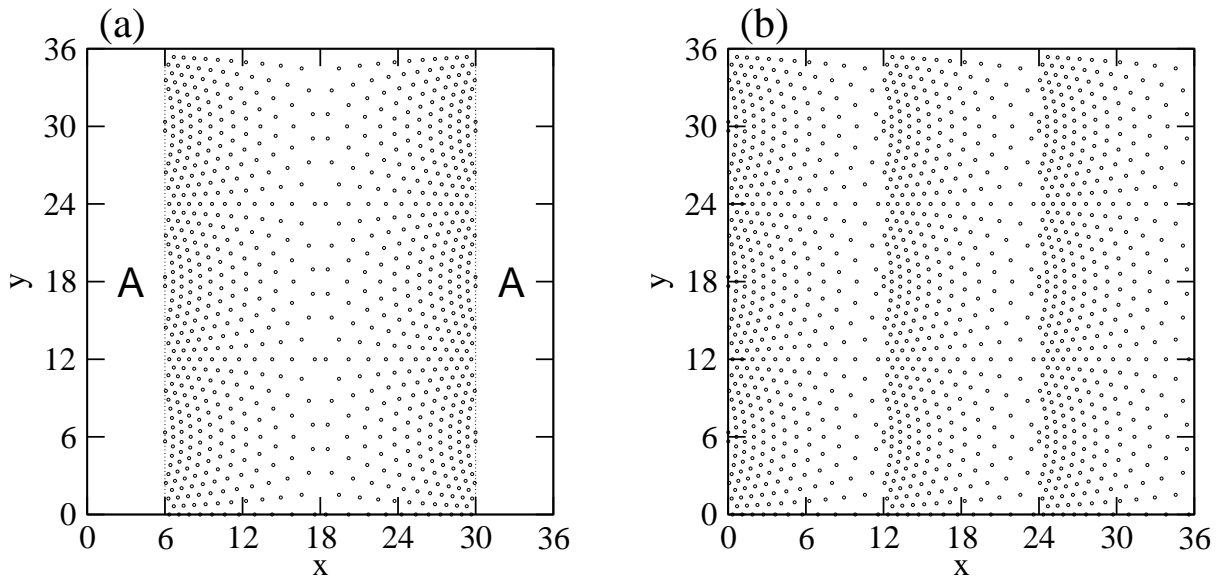}
\caption{ 
Pinning geometries for both types of simulations, 
illustrated using the cHex array.  
In each case, the sample is periodic in the $x$ and $y$ directions.
(a) Magnetization simulation geometry. The pinned region 
representing the superconducting sample 
extends from $x=6\lambda$ to $x=30\lambda$.  
Two copies of a gradient pinning array are placed in the 
pinned region, with their sparse sides facing each other. 
Vortices and antivortices are added to the unpinned outer
regions labeled `A', which represent an applied external field. 
(b) Transport simulation geometry. The pinned region spans 
the entire simulation box. 
Three copies of a gradient pinning array are placed in the box, 
all facing the same direction. Vortices are placed randomly and
annealed before the simulation begins; 
during the simulation, a steadily increasing drive is applied.
}
\label{fig:chexgeom}
\end{figure*}

We employ two types of simulation geometries in this work.
The first is a flux gradient-driven geometry \cite{41,46,49,50,51}
where we apply a slowly varying magnetic field to a 
superconductor with pinning sites  
and calculate the resulting magnetization loops. 
Specifically, we consider a two-dimensional
($x-y$) slice of a bulk superconductor
subjected to an applied magnetic field ${\bf H} = H{\bf \hat z}$. 
We treat the resulting vortices in the material as perfectly stiff 
and use the London limit in which the vortices are 
assumed to be pointlike in the $x-y$ plane.
We measure all lengths in terms of the penetration depth $\lambda$. 
Our system is a square of side $L=36\lambda$ 
with periodic boundary conditions 
in both the $x$ and $y$-directions. 
As shown in Fig.~\ref{fig:chexgeom}(a), 
the pinning is added in a region 
extending from $x = 6\lambda$ to $30\lambda$
which represents the superconducting sample.
Vortices and antivortices are added in the 
unpinned external regions labelled `A' 
to the left and right of the sample; 
the vortex density in these regions represents the applied external 
magnetic field.
For pinning arrays with a gradient, we place two copies of the array 
in the sample region, oriented so that their dense sides face the 
external regions, while their sparse sides face each other. 
The geometry and system size described here 
were previously shown to be sufficiently large to 
accurately model both magnetization curves and vortex ordering 
for conformal crystal pinning arrays \cite{41}, 
periodic pinning arrays \cite{41,50,51},
and random pinning \cite{41,46,49}.  

At a given time during the simulation, the system contains a total
of $N_v$ vortices and antivortices. As the external field is changed
over the course of the simulation, $N_v$ varies.
The equation of motion for vortex $i$ is given by
\begin{equation}
\eta \frac{ d {\bf R}_{i}}{ dt}  = {\bf F}^{vv}_{i} + {\bf F}^{p}_{i}  .
\end{equation}
Here ${\bf R}_{i}$ is the location of the vortex 
and $\eta$ is the damping constant which we set to unity.  
% vv force
The vortex-vortex interaction force is
\begin{equation*}
{\bf F}^{vv}_{i} = \sum_{j \neq i} s_{i}s_{j}F_{0}K_{1}
(R_{ij}/\lambda){\hat{\bf R}}_{ij}
\end{equation*} 
where $F_{0} = \phi_{0}^{2}/2\pi\mu_{0}\lambda^3$,
$\phi_{0}=h/2e$ is the elementary flux quantum, $\mu_{0}$ is the
permittivity of free space, 
$K_{1}$ is the modified Bessel function, 
$R_{ij} = |{\bf R}_{i} - {\bf R}_{j}|$, 
and ${\hat {\bf R}}_{ij} = ({\bf R}_{i} - {\bf R}_{j})/R_{ij}$.
We measure all forces in terms of $F_0$. 
For computational efficiency, 
we truncate the interaction force beyond $R_{ij}=6\lambda$, 
which is a good approximation since $K_{1}(x)$ falls off 
exponentially for large $x$.
Vortices repel each other, as do antivortices, while a vortex 
attracts an antivortex; to represent this interaction, we use
the sign prefactor $s_{i}$ which is $+1$ for a vortex and $-1$ for an
antivortex. 
If a vortex and an antivortex approach each other
within a distance smaller than
$0.3\lambda$, they are both removed from the system 
to simulate an annihilation event. 
Most of our simulations involve only vortices; we use antivortices
only to generate full magnetization loops, described below.
% pin force
The pinning arises from $N_{p}$ non-overlapping traps each represented
by a truncated parabolic pinning potential, so that the pinning force 
is given by 
\begin{equation*}
{\bf F}_{i}^{p}=\sum_{k=1}^{N_{p}}(F_pR_{ik}^p/R_{p}) 
\Theta\big((R_{p}-R_{ik}^p)/\lambda\big){\hat{\bf R}}_{ik}^p ,
\end{equation*}
where $R_k^p$ is the location of pinning site $k$, 
$R_{ik}^p=|{\bf R}_i-{\bf R}_k^p|$, 
$\hat{\bf R}_{ik}^p=({\bf R}_i-{\bf R}_k^p)/R_{ik}^p$, 
$\Theta$ is the Heaviside step function, 
the pinning radius $R_p$ is set to $0.12\lambda$, 
and $F_p$ specifies the maximum strength of the pinning force.

% mag sim
To construct a full magnetization loop, we begin with an empty 
sample and gradually add vortices to the unpinned regions 
marked `A' in Fig.~\ref{fig:chexgeom}(a). 
As the vortex density increases in the 
unpinned region, vortices are pushed into the sample due to the 
intervortex repulsion. We continue to add vortices until reaching 
the desired maximum value of the external magnetic field $H$; 
then we begin to add antivortices. As these annihilate with vortices, $H$ 
decreases back to zero and then becomes negative as the unpinned 
region fills with antivortices. Finally, we add vortices again to
bring $H$ back up to zero. As the simulation progresses, we
record the magnetization 
\begin{equation*}
M = -\frac{1}{4\pi V} \int dV\, (H-B)
\end{equation*}
where $V$ represents the sample area, $H$ is the average vortex 
density over the unpinned region, and $B$ is the (position-dependent)
vortex density in the sample region. 
According to the Bean critical state model \cite{52},
the width of the magnetization loop is 
proportional to the critical current $J_c$. 
This width can be determined using only the first quarter 
of the magnetization loop, 
and so in many of our simulations, we generate only a quarter loop
instead of a full loop by using vortices 
to raise the external field up to a maximum value and then halting
the simulation.

%trans sim
The second simulation geometry we consider is current-driven rather 
than flux gradient-driven, and is similar to that used in
previous studies of current-driven vortex
dynamics in pinned superconductors \cite{40,53,54}. 
We add
an additional force term ${\bf F}^{d}=F_{d}F_0\hat{n}$
to the vortex equation of motion in Eqn.~1 to 
represent the Lorentz force exerted by
an applied external current ${\bf J}$; 
here $\hat{n}$ is a constant unit vector in the $x-y$ plane. 
This force affects all vortices uniformly.
For the purposes of an initial anneal, we also add a term ${\bf F}_i^T$
representing thermal Langevin kicks to Eqn.~1, where ${\bf F}_i^T$ has
the properties
$\langle {\bf F}_i^T\rangle=0$ and 
$\langle {\bf F}^T_i(t) {\bf F}^T_j(t^\prime)\rangle=2\eta k_BT 
\delta_{ij}\delta(t-t^\prime)$.
We modify the system geometry to measure transport properties by 
removing the external unpinned regions and employing a pinning 
configuration composed of three copies of a given pinning array 
facing in the same direction and filling the entire simulation box, as
shown in Fig.~\ref{fig:chexgeom}(b).
To initialize the system we place
$N_v$ vortices in randomly chosen non-overlapping locations in the simulation
box and perform a simulated anneal from $F^T=3.0$ to $F^T=0$ while holding
$F_d=0$ before beginning our measurement.  
Throughout a transport simulation, $N_v$ is held constant in order
to model a constant magnetic field level.
We increase the magnitude 
of the current-induced driving force $F_d$ 
in small increments of size $\delta F_d=5 \times 10^{-4}$ and spend 
$500$ simulation time
steps at each value of $F_d$ in order to
avoid any transient behavior. 
We measure the average vortex velocity in the $x$ and $y$ 
directions as a function of drive: 
$\langle V_{x} \rangle = (1/N_{v})\sum^{N_{v}}_{i=1} {\bf v}_{i}\cdot\hat{x}$ and 
$\langle V_{y} \rangle = (1/N_{v})\sum^{N_{v}}_{i=1} {\bf v}_{i}\cdot\hat{y}$.
The vortex velocity is proportional to the voltage response, so 
the velocity-force curve we generate is comparable to
an experimentally measurable current-voltage curve.

\section{Magnetization}

\begin{figure}
\includegraphics[width=\columnwidth]{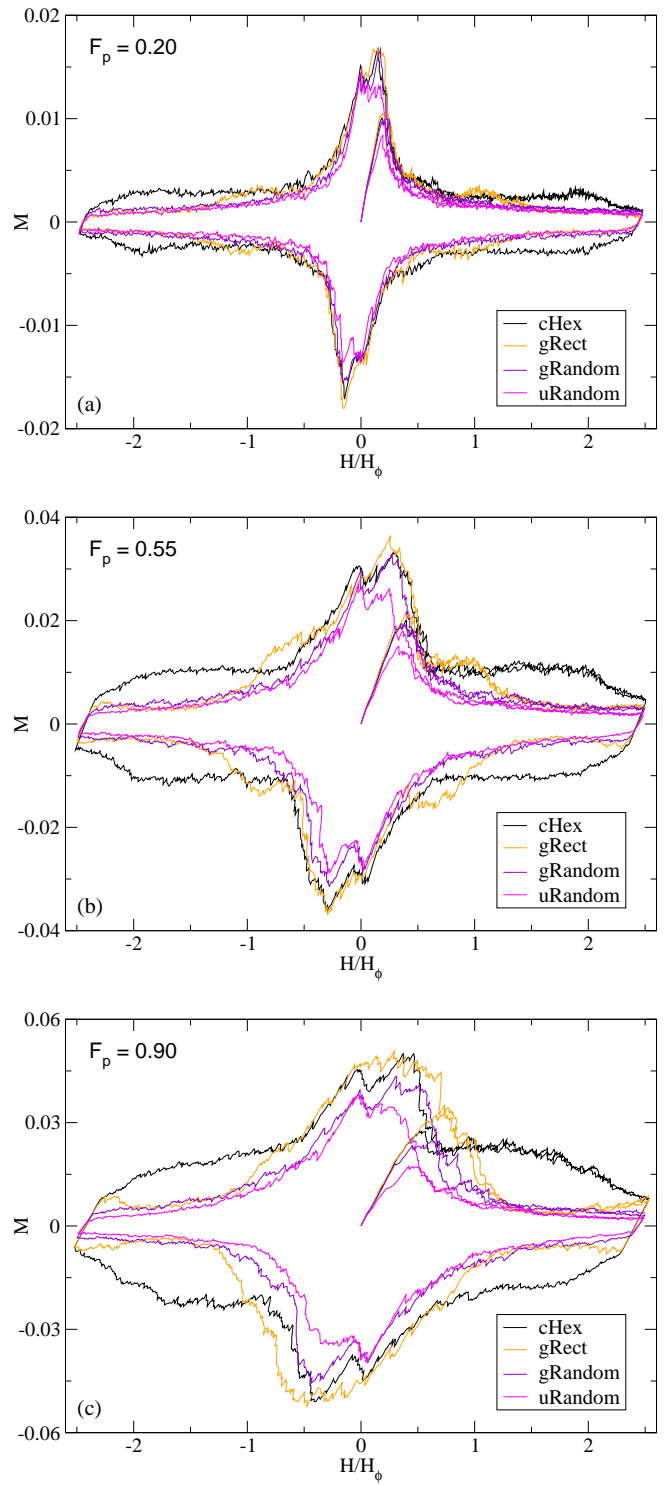}
\caption{ 
Full magnetization $M$ vs. external magnetic field 
$H/H_\phi$ loops 
for the conformal cHex 
array (outer loop), 
the non-conformal gRect and gRandom arrays (center loops), and uniform 
random pinning uRandom (inner loop), with
$F_p=$ 
(a) $0.2$, (b) $0.55$, and (c) $0.9$. 
The conformal array shows enhanced pinning over the 
other arrays in each case, and this is most noticeable
when $|H/H_{\phi}| > 1.0$. 
}
\label{fig:5}
\end{figure}

In Fig.~\ref{fig:5} 
we plot the full magnetization loops $M$ vs $H/H_{\phi}$,
where $H_{\phi}$ is the matching field at which 
there is one vortex per pinning
site, for the conformal pinning array 
cHex 
from Fig.~\ref{fig:transf}, 
the non-conformal gradient arrays gRect and gRandom 
from Fig.~\ref{fig:garray}, 
and a uniform random pinning array. 
For weak pinning sites with $F_p=0.2$,
Fig.~\ref{fig:5}(a) shows 
that the conformal array provides enhanced pinning 
compared to the other arrays over the entire range of field,
as indicated by the fact that the magnetization loop is widest for
the cHex array. 
The rectangular gradient array gRect and cHex have
similar values of $M$ up to $H/H_{\phi} = 1.0$, 
but for higher $H/H_{\phi}$ 
the conformal array is clearly the most effective at pinning the vortices.
The difference between the conformal and random arrays becomes more
pronounced when the pinning strength is increased, as indicated
for $F_p=0.55$ in Fig.~\ref{fig:5}(b).
The ordered gRect array 
shows slightly stronger pinning than cHex for $H/H_{\phi} <
1.0$, but falls off dramatically above that value.  
These effects are even more pronounced for $F_{p} = 0.9$,
shown in Fig.~\ref{fig:5}(c).
The difference between the uniform random and random gradient arrays
increases with increasing $F_p$, but the mere addition of a gradient is
not sufficient to give magnetization values that are anywhere close to as
large as those obtained with the cHex array, particularly for higher
values of $H/H_\phi$.
The effectiveness of the conformal array persists up to a field 
$H/H_\phi=2.0$ before beginning to decline, and 
the critical current for cHex remains higher than that of gRect
up to $H/H_\phi=2.5$. Above $H/H_\phi=1.75$ a new peak in $M$ begins
to emerge for the gRect sample for strong pinning; this feature is
described in more detail in Section \ref{sec:grectcomm}.
We note
that experiments in thin-film superconductors 
with through-hole pinning sites should exhibit strong pinning
behavior equivalent to large values of $F_p$ \cite{43}.
These results show the superiority of conformal gradient pinning 
over other types of gradient pinning, and also over random pinning
of any sort. 
In Fig.~4(d) of Ref.~\onlinecite{41} we have already showed
enhanced pinning for a conformal array 
compared to uniform periodic arrays over a broad range of fields, 
apart from certain narrow field ranges corresponding to matching conditions. 

\begin{figure}
\includegraphics[width=\columnwidth]{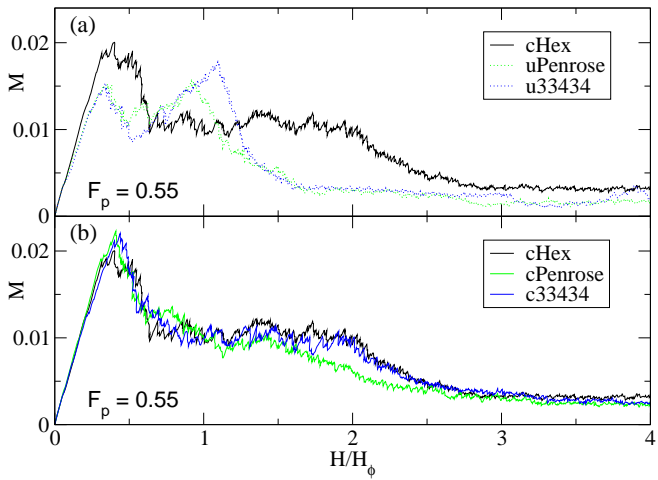}
\caption{ 
Magnetization $M$ vs $H/H_\phi$ for cHex array (top right) 
compared to various arrays with 
two length scales in samples with $F_p=0.55$.  
(a) Uniform 33434 (bottom right) and uniform Penrose
(center right); (b) conformal 33434 (center right) and 
conformal Penrose (bottom right).
}
\label{fig:pen55}
\end{figure}

\begin{figure}
\includegraphics[width=\columnwidth]{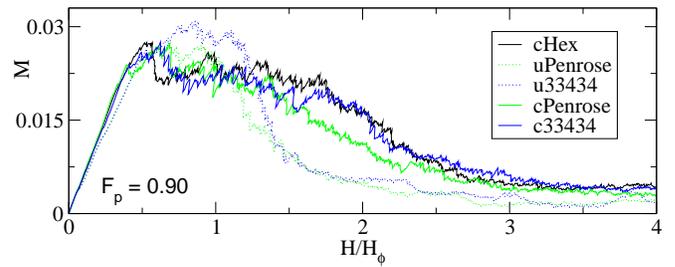}
\caption{ 
Magnetization $M$ vs $H/H_\phi$ for cHex array (top right)
compared to various arrays with 
two length scales in samples with $F_p=0.90$:
uniform 33434 (lower left), 
uniform Penrose (lower right), 
conformal 33434 (upper middle right), and conformal Penrose (center right).
}
\label{fig:pen90}
\end{figure}

We next turn to Penrose pinning arrays, since these have been claimed 
\cite{34,39} to also provide broadly effective pinning. 
The previous studies suffered from several drawbacks, 
such as a small sample and array size which diminishes any actual effect 
of the array's structure upon pinning effectiveness, the use of 
fractional pin occupancy as a proxy for critical current, and 
most importantly 
a lack of comparisons to other pinning structures besides triangular. 
Here, we compare a Penrose array generated from a tiling with 
approximate 5-fold symmetry, composed of thin and thick rhombs, 
to a conformal array. The magnetization curves for 
the uniform Penrose array with $F_{p}=0.55$ 
in Fig.~\ref{fig:pen55}(a) show a matching peak near 
$H/H_\phi=1.0$, similar to that found for
a uniform periodic array.  Away from the matching field, the 
magnetization drops off rapidly, again similar to the behavior of
a periodic array. 
The magnetization of the cHex array is lower 
than that of the uPenrose array around the matching 
field, but is higher everywhere else. In particular, the cHex array gives
a large enhancement of the magnetization up to and slightly beyond
$H/H_\phi=2.0$.

These results suggest that the quasiperiodic nature of a 
Penrose array is relatively unimportant as far as pinning effectiveness 
is concerned, and in fact its behavior is comparable to that of
strictly periodic pinning structures.  Since the Penrose tiling employs two 
different types of tiles giving rise to multiple length scales
in the pinning structure, we 
use as a comparison a periodic tiling known as Archimedean 33434, 
which also uses two types of tiles, and which has been studied in 
the context of vortex pinning \cite{51,archi}.
In Fig.~\ref{fig:pen55}(a) we show that the magnetization of the uniform
33434 array
is broadly similar to that of the 
uniform Penrose array, with a sharp matching peak in $M$ 
at $H/H_\phi=1$ followed 
by a rapid falloff. 

Next, in order to examine the role of local order in conformal arrays, 
we consider the conformally transformed versions of the Penrose and 
Archimedean arrays which were shown in Fig.~\ref{fig:carray}.  We plot
magnetization curves for these arrays in Fig.~\ref{fig:pen55}(b).
Two behaviors are apparent. 
First, comparing the uniform arrays 
in Fig.~\ref{fig:pen55}(a) to their conformally transformed 
versions in Fig.~\ref{fig:pen55}(b), we see that the
conformal transformation improves an array's performance 
over a broad range of field at the cost of 
eliminating the original matching peak at $H/H_\phi$. 
Second, we find from Fig.~\ref{fig:pen55}(b) that even though
vortices naturally arrange themselves in a hexagonal lattice in the absence 
of pinning, local hexagonal ordering is {\it not} necessary 
to obtain an effective conformal array: the c33434 array works as well,
and the cPenrose array almost as well, as the cHex array at enhancing pinning.

To verify that these results are robust, we repeat our measurements in
samples with an increased pinning 
strength of $F_p=0.90$.  
As shown in Fig.~\ref{fig:pen90}, the magnetization curves 
retain all of the behaviors described above. In particular, we confirm that 
conformal arrays work better than uniform arrays in each case, and that their 
performance does not depend sensitively on the type of local order present
in the original untransformed array. 
To further test this, a future study could compare cHex and cSquare 
arrays in magnetization loops which extend up to the third matching field, 
since a uniform square array has a commensuration peak at the second 
matching field but not the third, while the uniform triangular array 
has the opposite response. 

\section{Local Commensuration Effect in Conformal Pinning Arrays} 
\label{sec:localcomm}

\begin{figure}
\includegraphics[width=\columnwidth]{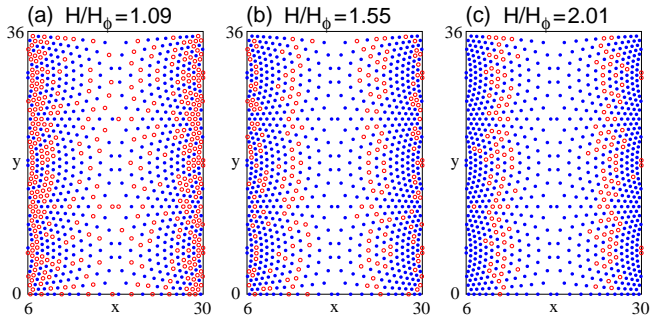}
\caption{ 
Occupied and empty pinning sites in a cHex array with $F_{p}=0.55$, 
at $H/H_{\phi}$= (a) 1.09, (b) 1.55, and (c) 2.01. 
Filled circles (blue): occupied pins; open circles (red): unoccupied pins.
Pin sizes have been exaggerated for clarity. 
The occupied pinning sites form vertical bands that move toward 
the sample edges with increasing field.
}
\label{fig:pinoccpix}
\end{figure}

\begin{figure}
\includegraphics[width=\columnwidth]{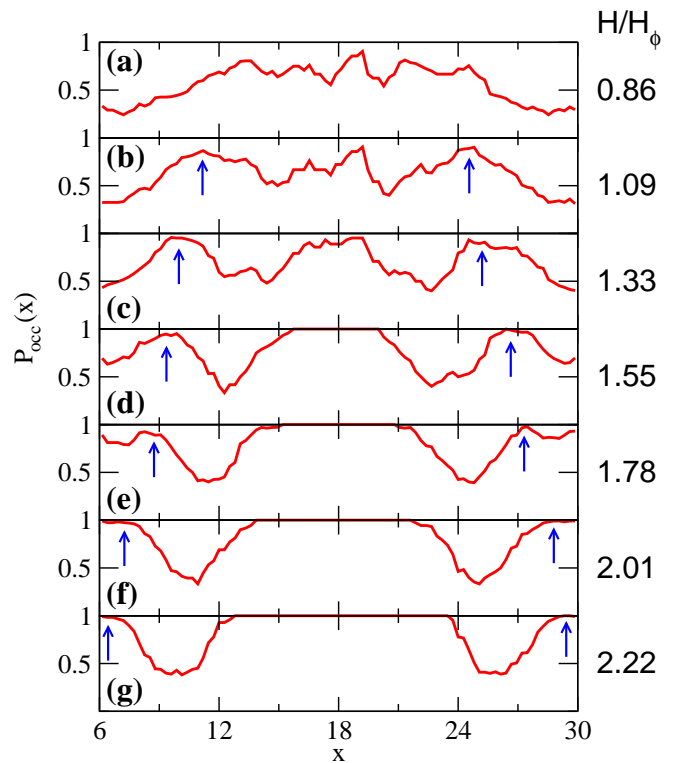}
\caption{ 
The spatial distribution of the fraction of occupied pins 
$P_{\mathrm{occ}}(x)$,  
averaged over the $y$-direction, versus $x$ position
for the cHex array with $F_{p} = 0.55$ at $H/H_{\phi}$=
(a) 0.86, (b) 1.09, (c) 1.33, (d) 1.55, (e) 1.78, (f) 2.01, and (g) 2.22.
The arrows in panels (b-g) highlight local
regions of very high pin occupancy, indicating that the vortex
lattice is locally commensurate with the pinning in these regions.
As $H/H_{\phi}$ increases, 
the matching regions move toward the sample edges where the pinning density 
is greater.   At the extreme edges of the
sample the maximum local pinning density is $\rho_p^{\rm loc}=2.0/\lambda^2$. 
}
\label{fig:9}
\end{figure}

In order to better understand 
how the structure of conformal arrays contributes to their 
enhanced pinning effectiveness 
over a broad range of field, 
and also what determines the limits of this range, 
we image the distribution of occupied pins as the field is increased. 
We take as our model system the cHex array with $F_{p}=0.55$.
In Fig.~\ref{fig:pinoccpix} we show three snapshots of the occupied
pinning sites at different field values, where we find 
vertical bands in which all the pinning sites are filled.
The location of these bands moves outwards towards the 
left and right edges of the sample as the field increases.  
This is seen more clearly 
in Fig.~\ref{fig:9}, where we plot the
fraction of occupied pins $P_{\mathrm{occ}}(x)$ 
averaged over the vertical direction $y$ 
as a function of horizontal position $x$
for several different values of $H/H_\phi$.
Here $x=18$ denotes the center of the pinned region, while the 
sample edges 
are at $x=6$ and $x=30$. 
For $H/H_{\phi} = 1.09$ in Fig.~\ref{fig:9}(b), 
we highlight two occupancy peaks
in $P_{\mathrm{occ}}(x)$ near $x = 11$ and $x = 25$.  
These peaks correspond to the bands mentioned above, 
and signify areas where the vortex lattice {\it locally} 
matches with the pinning.  This matching is distinct from
the usual matching effect 
in a uniform pinning array which is global and occurs throughout 
the entire array.  
As $H/H_{\phi}$ increases, Figs.~\ref{fig:9}(c-g) show that the
occupancy 
peaks move toward the sample edges. 
This occurs because 
at matching, the vortex density must match the pinning density; 
thus, as the vortex density increases, 
the matching region must shift to the denser 
pinning regions that are closer to the edges of the sample.
We also observe
dips in $P_{\mathrm{occ}}(x)$ adjacent to the peaks, 
on the sides closer to the
center of the sample.  The dips arise
due to the mismatch between vortices and pins once the commensuration
peak has passed, 
with interstitial vortices forcing vortices out of the pinning sites.  
The migration of the local matching peaks
continues as the field increases until the 
peaks reach the maximally dense pinning regions at the sample edges.
For the conformal arrays we use in this work, the maximum local pinning
density is $\rho_p^{\rm loc}=2.0$, 
so we expect that when the vortex density $\rho_v$
exceeds this value by a non-negligible amount, local matching 
will no longer be possible. 
In Fig.~\ref{fig:9}(g) where the external field has increased to 
$H/H_\phi=2.22$, 
the matching regions have reached the edges of the sample 
and can not move any further.  Figure \ref{fig:pen55} shows that
the effectiveness of the pinning of the
cHex array falls off above this field value.

\begin{figure}
\includegraphics[width=\columnwidth]{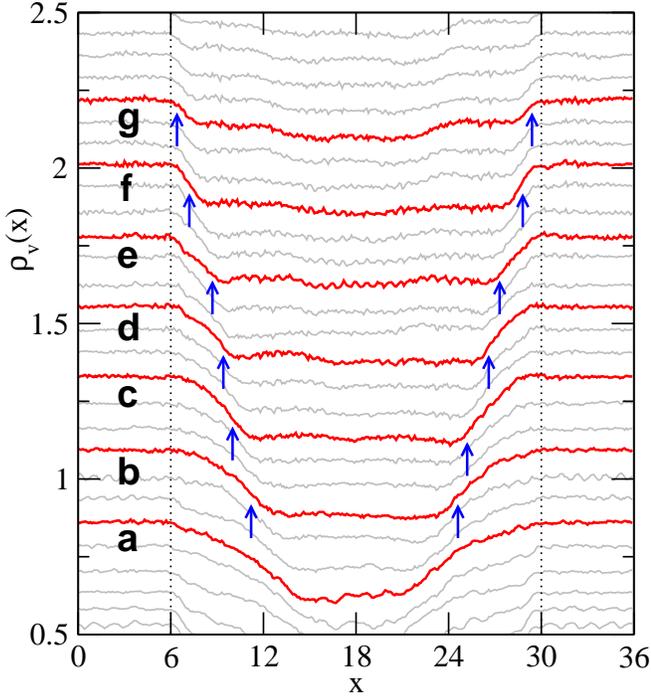}
\caption{
Vortex density $\rho_v$ vs $x$ position for the cHex array with 
$F_{p}=0.55$ as the external field is gradually increased. Dotted lines
indicate the sample edges; the sample extends from $x=6\lambda$ to
$x=30\lambda$ as shown in Fig.~\ref{fig:chexgeom}(a).
The value of $H/H_\phi$ for each curve is indicated by the average $y$
value in the external regions $x<6\lambda$ and $x>30\lambda$.
The lowest curve has $H/H_\phi=0.52$, 
the highest curve has $H/H_\phi=2.50$, and
there is an interval of 100 vortices, or 
$\delta H/H_\phi \approx 0.077$, between illustrated curves.
Profiles in red correspond to the magnetic field values shown in 
Fig.~\ref{fig:9}; the letters indicate the corresponding panel 
in Fig.~\ref{fig:9}.  
Blue arrows point to the same locations as the arrows in 
Fig.~\ref{fig:9}.
}
\label{fig:chexprof}
\end{figure}

\begin{figure*}
\includegraphics[width=7in]{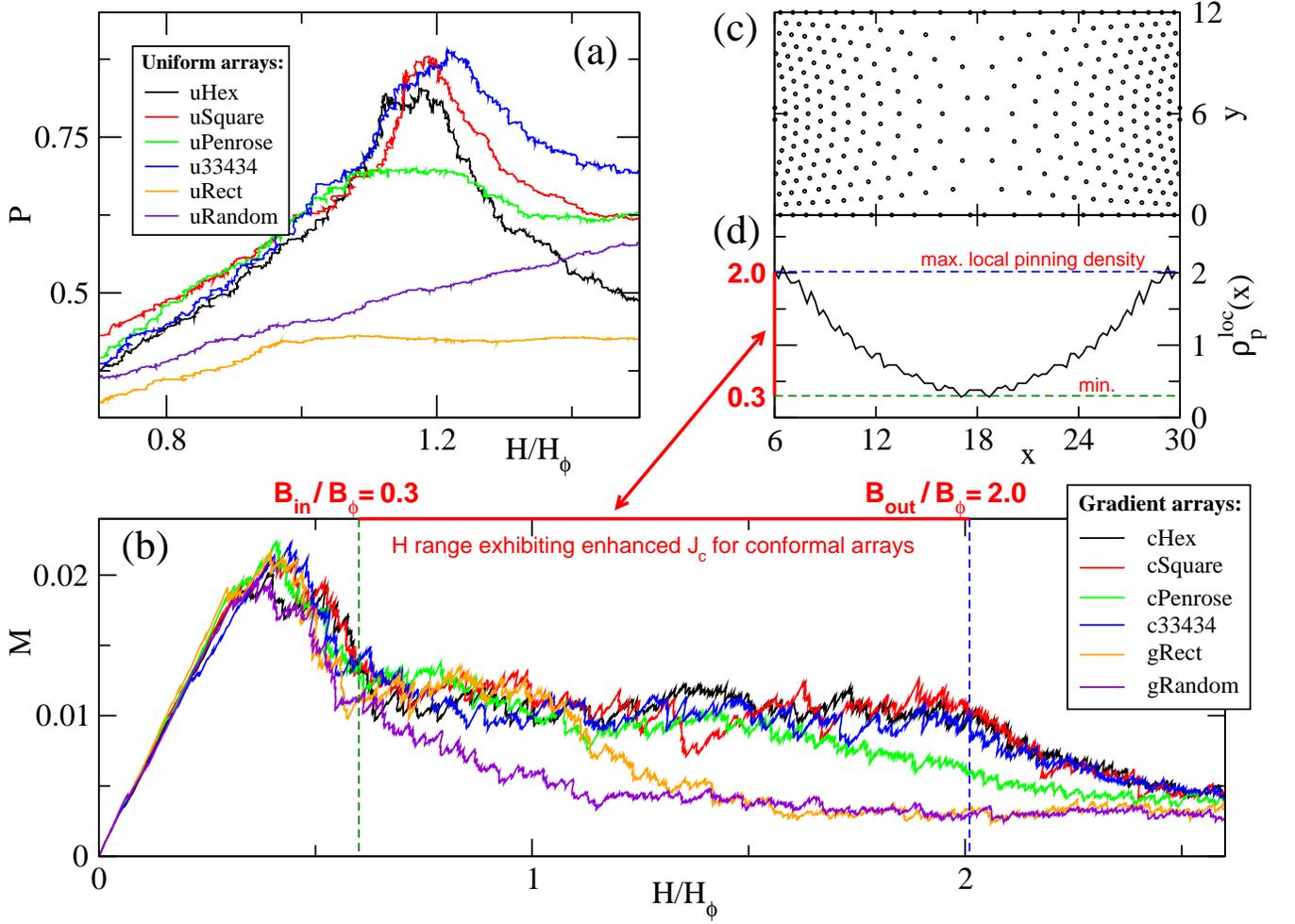}
\caption{
Local commensuration effects in conformal pinning arrays. 
(a) Spatially averaged fraction of occupied pins $P$ vs $H/H_\phi$ 
in {\it uniform} arrays of 
(central peak, from top to bottom) 
hexagonal, square, Penrose, 33434, rectangular, and random type.
Peaks in $P$ just above $H/H_\phi=1.0$ 
show that the hexagonal, square, Penrose, and 33434 arrays 
exhibit a robust matching effect. 
(b) Spatial configuration of the pinning sites in a cHex array. 
(c) Corresponding local pinning density $\rho_p^{\rm loc}(x)$ 
for the same (cHex) array.
The local pinning densities for the other conformal arrays are nearly
identical to what is shown here.  
The region between the dotted lines indicates the
range of $\rho_p^{\rm loc}$ values that exist somewhere inside the array.
(d) Magnetization curves $M$ vs $H/H_\phi$ for gradient pinning arrays
(from upper left to lower left): 
cHex, cSquare, cPenrose, c33434, gRect, and gRandom. 
For the conformal arrays, the magnetization is enhanced whenever
the vortex lattice can locally
match with a vertical strip of the pinning array. 
The range of possible local densities allowing matching 
[highlighted region in panel (c)]
thus determines the range of applied fields 
[highlighted region in panel (d)] at which the conformal
arrays provide enhanced values of $M$.
The non-conformal gradient arrays do not 
enhance pinning by this mechanism. 
}
\label{fig:11}
\end{figure*}

The vertical bands of occupied pins act as walls hindering vortices 
from flowing between the outer region and the inner region. These
effective walls can maintain
a large difference in vortex densities between these regions,
giving a large value of $M$.
This is illustrated in Fig.~\ref{fig:chexprof}, 
where we plot the vortex density $\rho_v$ averaged over $y$ as a 
function of horizontal position $x$
as the field is increased. 
The field profiles 
have a double-slope, non-Bean-like shape as noted in 
previous work \cite{41}. 
The high-slope regions where the vortex density changes rapidly 
are precisely the regions in which we find bands of occupied pins. 
We show this by drawing arrows pointing to 
the labeled highlighted profiles in Fig.~\ref{fig:chexprof} 
at the same locations where they appear in 
the corresponding pin occupancy plots in Fig.~\ref{fig:9}; 
these arrows clearly track the high-slope profile regions. 
The high-slope regions move towards the edges 
of the sample as the field increases, and first touch the
edges at $H/H_\phi=2.0$.  
Above this field, the high-slope regions 
begin to disappear since local matching can no longer occur,
the vortex density equilibrates between the sample and the outside, 
and there is a corresponding drop in $M$ as seen for cHex
in Fig.~\ref{fig:pen55}.

In Fig.~\ref{fig:11} we summarize the key characteristics of 
conformal arrays which produce the enhanced pinning. 
Although the data presented in the above discussion of local 
commensuration was from the cHex array, the local
commensuration mechanism does 
not require hexagonal ordering; in fact, any local ordering 
that exhibits matching phenomena should be similarly effective. 
To show this, we first examine which array types are capable 
of exhibiting matching phenomena. 
Figure \ref{fig:11}(a) shows the 
spatially averaged fraction of occupied pinning sites $P$ vs
$H/H_{\phi}$ for {\it uniform} arrays of the 
hexagonal, square, Penrose, 33434 (Archimedean), rectangular,
and random type.
The hexagonal, square, and 33434 
arrays have a pronounced peak in $P_{\rm occ}$ just above
$H/H_{\phi} = 1.0$, indicating that they exhibit robust matching. 
The Penrose array shows weaker matching, 
while the random and the rectangular arrays do not exhibit matching at all.
In Fig.~\ref{fig:11}(b) we plot 
$M$ versus $H/H_\phi$ for the corresponding 
conformal or gradient arrays, cHex, cSquare, cPenrose, c33434, gRect, and
gRandom.  As expected from the arguments above, we find
that the conformal versions of the 
hexagonal, square, and 33434 arrays all show highly enhanced pinning 
up to $H/H_{\phi} = 2.0$, 
resulting from robust local matching. 
The conformal Penrose array shows a weaker pinning enhancement since
the commensuration effects in the original uniform Penrose array 
are weaker.
The rectangular and random gradient arrays give much lower
magnetization values than the other arrays
because they are not conformal and do not exhibit a local matching
mechanism. 
We conclude that for any
uniform array with a well-defined matching condition,
the corresponding
conformal array will exhibit enhanced pinning over
a broad range of fields.
For the conformal arrays we consider, this field range is
$0.6 < H/H_{\phi} < 2.0$, and it is determined by
the local pinning density $\rho_{p}^{\rm loc}(x)$ 
which we plot in Fig.~\ref{fig:11}(d) for the conformal hex 
lattice shown in Fig.~\ref{fig:11}(c).
At the minimum external field $H/H_{\phi} = 0.6$, the local 
matching occurs at the minimum density region in the center 
of the sample, where $\rho_{p}^{\rm loc} = 0.3$. 
The local matching moves 
toward the sample edges as the field increases, 
until for $H/H_\phi=2.0$ the matching reaches the sample edges where
the pinning density is also $\rho_p^{\rm loc}=2.0.$
Future experimental studies can determine the maximum steepness 
of the pinning gradient 
which will still permit local matching to occur and be effective 
at enhancing the pinning.

\section{Rectangular Gradient Array and Commensuration Effects}
\label{sec:grectcomm}

\begin{figure}
\includegraphics[width=\columnwidth]{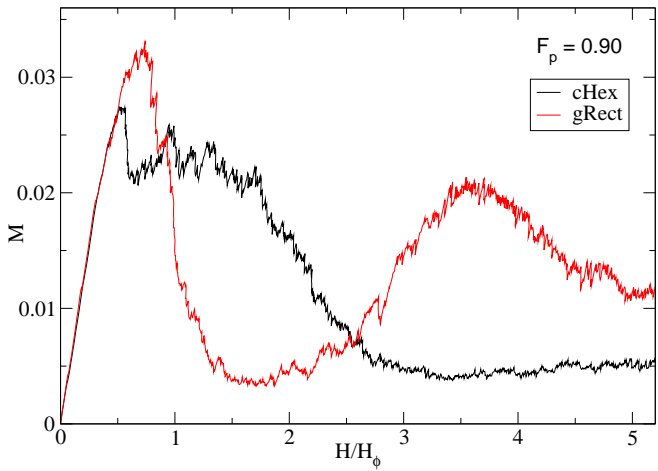}
\caption{ 
$M$ vs $H/H_{\phi}$ for cHex (lower right) and gRect (upper right) samples
with 
$F_{p} = 0.9$. 
Here the gRect array shows a commensuration peak near
$H/H_{\phi} = 3.5$, where the magnetization is significantly enhanced compared
to the cHex array.
}
\label{fig:12}
\end{figure}

We next consider higher-field ordering and commensurability
effects for the rectangular gradient or gRect pinning array, 
which were briefly noted in Fig.~\ref{fig:5}(c).  
In Fig.~\ref{fig:12} we plot $M$ versus $H/H_\phi$ for both
the cHex and gRect arrays with $F_{p} = 0.9$ for fields up
to $H/H_{\phi} = 5.2$. 
Here the effectiveness of the pinning for the conformal 
array clearly diminishes
for fields above $H/H_{\phi} \approx 2.0$.  
The cHex array gives enhanced pinning compared to the gRect array
for $1.0 \leq H/H_\phi \leq 2.5$; 
however, above this field range, $M$ for the gRect array begins to increase
dramatically and rises well above $M$ for the cHex array, forming
a broad peak centered at $H/H_{\phi} = 3.5$.

\begin{figure}
\includegraphics[width=\columnwidth]{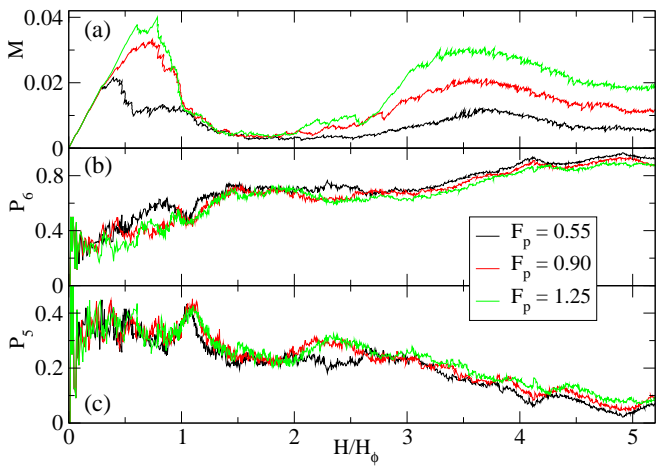}
\caption{ 
Results for the rectangular gradient array illustrated in 
Fig.~\ref{fig:garray}(a).
(a) $M$ vs $H/H_{\phi}$ for $F_{p} = 1.25$ (upper curve), 
$0.9$ (middle curve) and $0.55$ (lowest curve). 
(b) The corresponding fraction of 
sixfold-coordinated vortices $P_{6}$
vs $H/H_{\phi}$ for $F_p=1.25$ (lower right), $0.9$ (middle right), 
and $0.55$ (upper right).
There is a small peak in $P_{6}$ near $H/H_{\phi} = 4.0$.
(c) The corresponding fraction of
fivefold-coordinated vortices $P_{5}$ vs $H/H_\phi$
for $F_p=1.25$ (upper right), $0.9$ (middle right), and $0.55$ (lower right). 
}
\label{fig:13}
\end{figure}

In Fig.~\ref{fig:13}(a),  
we plot $M$ versus $H/H_{\phi}$ for rectangular gradient arrays with
$F_{p} = 0.55$, 0.9, and $1.25$,
showing that the broad peak in $M$ is a robust feature.  
In Fig.~\ref{fig:13}(b) we plot the
corresponding fraction of sixfold coordinated vortices
$P_{6}$ versus $H/H_{\phi}$, where $P_6=(1/N_{vs})\sum_{i=1}^{N_{vs}}\delta(z_i-6)$.
Here $N_{vs}$ is the number of vortices within the pinned region, 
the coordination number $z_i$ of vortex $i$ is obtained from a
Voronoi tesselation, and
for a triangular lattice $P_{6} = 1.0$. 
Figure~\ref{fig:13}(c) shows the corresponding 
fraction of fivefold coordinated vortices $P_{5}$ versus $H/H_{\phi}$. 
Due to the gradient in the sample, we generally find $P_7 \ll P_5$ since
vortices are moving into less dense areas, so we do not show $P_7$.
In Fig.~\ref{fig:13}(a), $M$ increases with increasing $F_{p}$ 
when $H/H_\phi\leq 1.0$, 
drops to a low value independent of $F_{p}$ 
for $1.0 < H/H_{\phi} < 2.0$, 
and increases with increasing $F_{p}$ again
for $2.0 \leq H/H_\phi$.  
For $H/H_{\phi} \lesssim 1.0$, 
vortices that enter the system tend to get captured by pinning sites, 
and the capture process is more efficient when $F_p$ is higher.
As in all computed magnetization curves irrespective of pinning geometry,
the initial change in slope of $M$ from positive to negative occurs
at the field at which the vortices first reach the center of the
sample.
Above this field, the vortices begin to enter the interstitial
regions of the rectangular gradient array,
where they encounter one-dimensional easy-flow channels oriented along
the $y$ direction, perpendicular to the gradient.
The interstitial vortices, although unpinned, 
experience an effective pinning force due to caging by
the surrounding pinned vortices.
This effective force is not affected 
by an increase in $F_{p}$, which only traps the pinned 
vortices more firmly. 
Thus, in the field range
$1.0 < H/H_{\phi} < 2.0$ where interstitial pinning is dominant,
we find that the value of $M$ is nearly independent of $F_p$. 
There is a dip in $P_6$ in 
Fig.~\ref{fig:13}(b) just above $H/H_\phi=1.0$,
with a corresponding peak in $P_{5}$ in Fig.~\ref{fig:13}(c), 
at the transition to the interstitial vortex regime.

\begin{figure}
\includegraphics[width=\columnwidth]{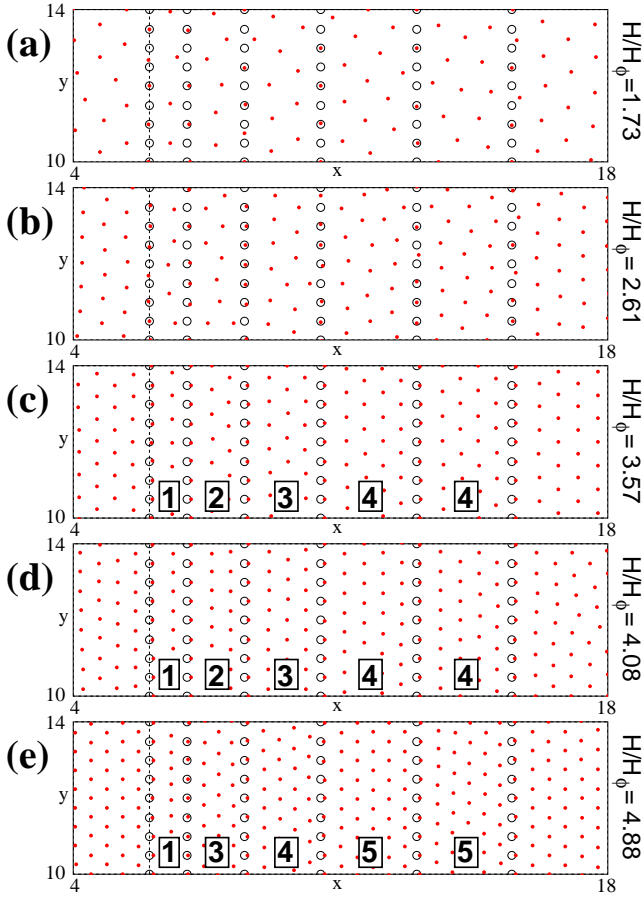}
\caption{ 
Ordered and disordered vortex states in rectangular gradient array 
with $F_{p}=0.90$ at various values of $H/H_\phi$.
Filled circles: vortices; open circles: pinning sites.
The dashed line on the left indicates the edge of the sample. 
(a) At $H/H_{\phi} = 1.73$, $M$ reaches a minimum
value and the vortices are disordered.
(b) At $H/H_\phi=2.61$, $M$ increases and the number of occupied
pins increases.
(c) At $H/H_\phi=3.57$, there is an ordered state 
corresponding to the peak in $M$ 
in Fig.~\ref{fig:13}(a). 
The numbers indicate the number of ordered vortex columns 
between columns of pinning sites. 
(d) At $H/H_\phi=4.08$, we observe an ordered state at 
the $P_{6}$ peak  
in Fig.~\ref{fig:13}(b). 
(e) At $H/H_\phi=4.88$ we illustrate the ordered state 
at the second $P_{6}$ peak 
in Fig.~\ref{fig:13}(b). 
}
\label{fig:14}
\end{figure}

For $H/H_{\phi} > 3.25$, Fig.~\ref{fig:13} shows that
$P_{6}$ begins to increase as $M$ rises to a broad peak
centered near $H/H_\phi=3.75$. 
There is a small
peak in $P_{6}$ near $H/H_{\phi} = 4.1$ followed by a small
decrease in the region where $M$ is also decreasing.
The broad peak in $M$ at these higher fields
is due to a commensuration effect 
associated with
the formation of ordered vortex chain
structures, as illustrated in Fig.~\ref{fig:14} where we show the
vortex configurations in the gRect array for different field values.
Figure~\ref{fig:14}(a) shows the configuration at $H/H_{\phi}=1.74$, 
where there is a minimum in $M$ in Fig.~\ref{fig:13}(a). 
Here the vortices are largely disordered, and substantial 
depinning has occurred as indicated by the large fraction of empty
pinning sites. 
At $H/H_{\phi} = 2.61$ in Fig.~\ref{fig:14}(b), 
$M$ begins to increase and the pinning sites fill, but there is
still no clear ordering present. 
When $M$ reaches its maximum at $H/H_{\phi} = 3.57$,
Fig.~\ref{fig:14}(c) shows that 
the vortex configuration within the pinning array is now ordered.
To highlight this ordering, we mark the number of vortex chains
that fit between pinning columns in Fig.~\ref{fig:14}(c). 
Between the two leftmost columns, there is one chain of
vortices; between the next two columns, there are two chains 
with the vortices forming a zig-zag structure; 
between the next two columns, there are three chains forming
a local triangular lattice, and similarly in the next two gaps
there are four chains.
Similar structures, but with different numbers of vortex chains 
in between pinning columns, appear
in Fig.~\ref{fig:14}(d) at $H/H_{\phi} = 4.08$ 
where there is a peak in $P_{6}$ in Fig.~\ref{fig:13}(b),
as well as at $H/H_{\phi} = 4.88$ in Fig.~\ref{fig:14}(e), 
where there is a second peak in $P_{6}$ in Fig.~\ref{fig:13}(b).

\section{Current-driven Results}

\begin{figure}
\includegraphics[width=\columnwidth]{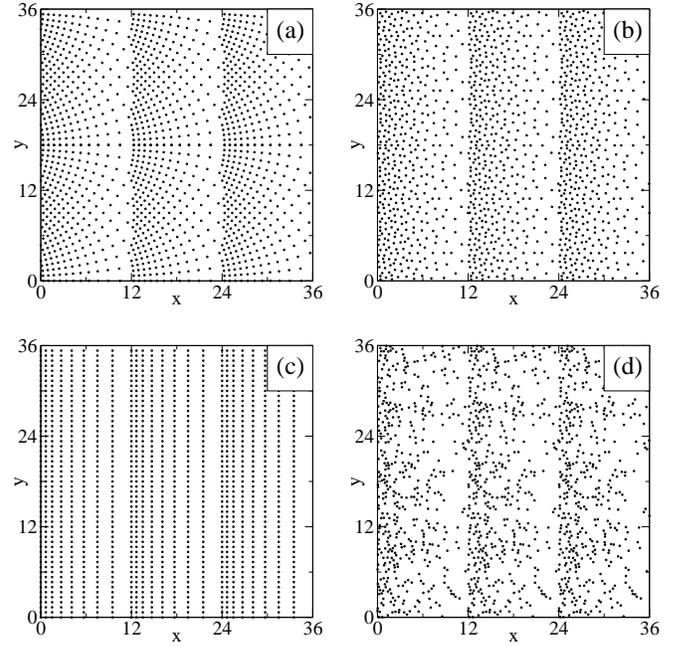}
\caption{
Pinning geometries for current-driven simulations with various 
pinning arrays: 
(a) cSquare; (b) cPenrose; (c) gRect; (d) gRandom. 
}
\label{fig:15}
\end{figure}

In the previous sections, we studied vortex ordering and critical states 
using quasistatic flux gradient-driven simulations. 
To examine the vortex dynamics, 
we use current-driven simulations in which a
steadily increasing drive
is uniformly applied to the vortices in the sample. 
We consider driving directions that are aligned with the gradient ($x$)
direction as well as drives oriented 
at some angle $\theta$ to this direction.
The equations of motion for the
current-driven simulations were described in Section III;
the sample geometry is illustrated in Fig.~\ref{fig:15}. 

\subsection{Drive Aligned with Gradient}

\begin{figure}
\includegraphics[width=\columnwidth]{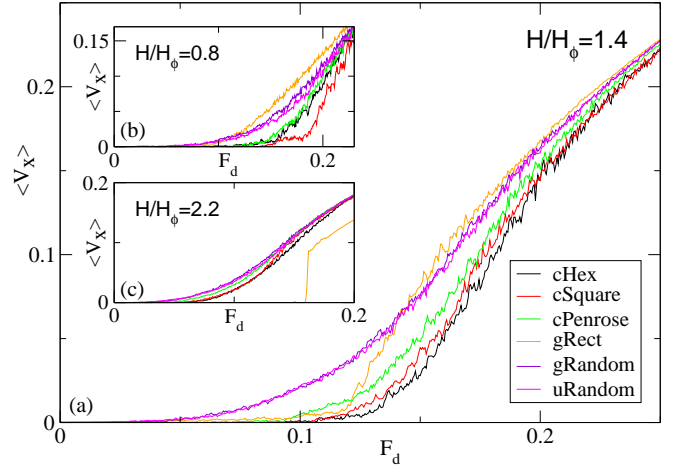}
\caption{ 
The average vortex velocity in the $x$-direction $\langle V_{x} \rangle$ 
vs $F_{d}$ for current-driven simulations of  
cHex, cSquare, cPenrose, gRect, gRandom, and uRandom transport arrays 
with $F_{p} = 0.55$ and a drive angle of $\theta=0$
at different vortex densities.  
(a) $H/H_{\phi} = 1.4$ for cHex, cSquare, cPenrose, uRandom, gRandom, 
and gRect,  
from lower right to upper right. 
(b) $H/H_{\phi} = 0.8$ for cSquare, cHex, cPenrose, uRandom, gRandom, 
and gRect, 
from lower right to upper right. 
(c) $H/H_{\phi} = 2.2$ for gRect,  
cHex, cSquare, cPenrose, uRandom, and gRandom,
from lower right to upper right. 
In general, the conformal arrays have a higher depinning threshold and lower 
vortex velocity than the non-conformal arrays. 
The gRect array exhibits a large depinning threshold at 
$H/H_{\phi} = 2.2$ due to a commensuration effect. 
}
\label{fig:16}
\end{figure}

We first consider driving vortices along the gradient in the
positive $x$-direction. 
We define the driving angle $\theta$ as the angle the driving 
direction makes with the positive $x$ axis, so that $\theta=0$
corresponds to driving along the gradient direction.
In Fig.~\ref{fig:16}(a) we plot $\langle V_{x} \rangle$ 
versus $F_{d}$ for systems with 
$F_{p} = 0.55$, $\theta=0$, and $H/H_{\phi} = 1.4$; 
we consider the cHex array shown in Fig.~\ref{fig:chexgeom}(b), 
the cSquare, cPenrose, gRect, and gRandom arrays 
shown in Fig.~\ref{fig:15} 
and a uniform random array lacking any gradient. 
The depinning force $F_c$, defined as
the magnitude of the drive above which
$\langle V_{x} \rangle \geq 10^{-4}$, is significantly larger 
in the conformal arrays than in the non-conformal arrays.
Additionally, in the moving phase the vortices travel more slowly 
through conformal arrays than through the nonconformal
arrays over a substantial range of $F_{d}$.  
Among the conformal arrays, the cHex array gives the highest $F_c$,
with $F_c$ for the cSquare array almost as high and 
slightly smaller for the cPenrose array.
At higher $F_{d}$ the curves start to come together as the
effectiveness of the pinning at the higher drives is washed out 
and $\langle V_{x} \rangle \approx F_{d}$. 
At $H/H_\phi=0.8$, as shown in Fig.~\ref{fig:16}(b),
$\langle V_x\rangle$ for the cSquare array drops below that of the
cHex array at higher drives, indicating enhanced pinning effectiveness
in the cSquare array,
while the curves for the cHex and cPenrose arrays are very similar.
The gRect array has the weakest pinning, followed by the 
random arrays with and without a gradient.

In Fig.~\ref{fig:16}(c) we illustrate the behavior at 
a higher field of $H/H_\phi=2.2$.
At this field, the gRect array exhibits a highly enhanced 
depinning force, followed by an abrupt transition to steady vortex 
motion. Inspection of the vortex configurations reveals that 
the vortices form an ordered vortex chain state of the type 
discussed in Section \ref{sec:grectcomm}, which locks them into 
place and prevents vortex flow. 
Because our gRect transport and magnetization 
arrays have somewhat different configurations, 
they form these ordered states at different field levels; 
however, the qualitative behavior of gRect remains the same 
during transport, with the array exhibiting enhanced performance 
only when it is able to form vortex chain states. 
The difference between 
the conformal and non-conformal random arrays 
is somewhat reduced at $H/H_\phi=2.2$, 
since it is 
above the cutoff field value of $H/H_\phi=2.0$ where the conformal 
arrays start to exhibit decreased pinning effectiveness. 
The conformal arrays 
still perform better by giving higher values of $F_c$ and lower values
of $\langle V_x\rangle$ in the moving phase than the random arrays.
The cSquare and cHex arrays have almost the same behavior, 
and are again superior to cPenrose. 

These results are completely consistent with the results from 
our magnetization simulations.  The transport simulations confirm 
that the pinning effectiveness is enhanced in conformal arrays over 
non-conformal ones, 
and the fact that the gRandom and uRandom arrays give nearly 
identical results indicates 
that it is not merely the gradient in the pinning that matters 
but also the local ordering. 
The precise type of local ordering is of lesser importance, 
although (as with magnetization) the quasiperiodic Penrose 
ordering is somewhat less effective at enhancing the
pinning than periodic orderings. 

\subsection{Driving in Different Directions} 

\begin{figure}
\includegraphics[width=\columnwidth]{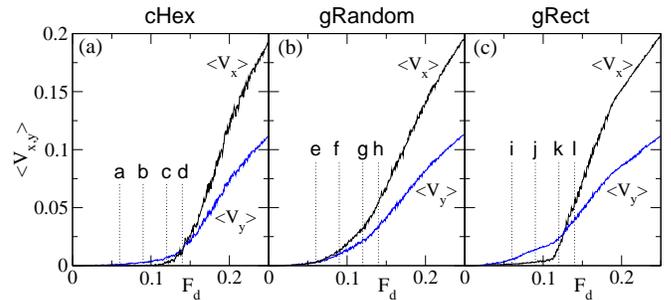}
\caption{ 
$\langle V_{x} \rangle$ and $\langle V_{y} \rangle$ 
vs $F_{d}$ for current-driven systems 
with $F_{p} = 0.55$ at $H/H_{\phi} = 1.4$,  
driven at
$\theta=30^{\circ}$ with respect to the positive $x$-axis. 
(a) cHex array; (b) gRandom array; (c) gRect array. 
The labeled current levels {\it a-l} 
correspond to the vortex trajectories illustrated in Fig.~\ref{fig:18}.
}
\label{fig:17}
\end{figure}

\begin{figure*}
\includegraphics[width=6.5in]{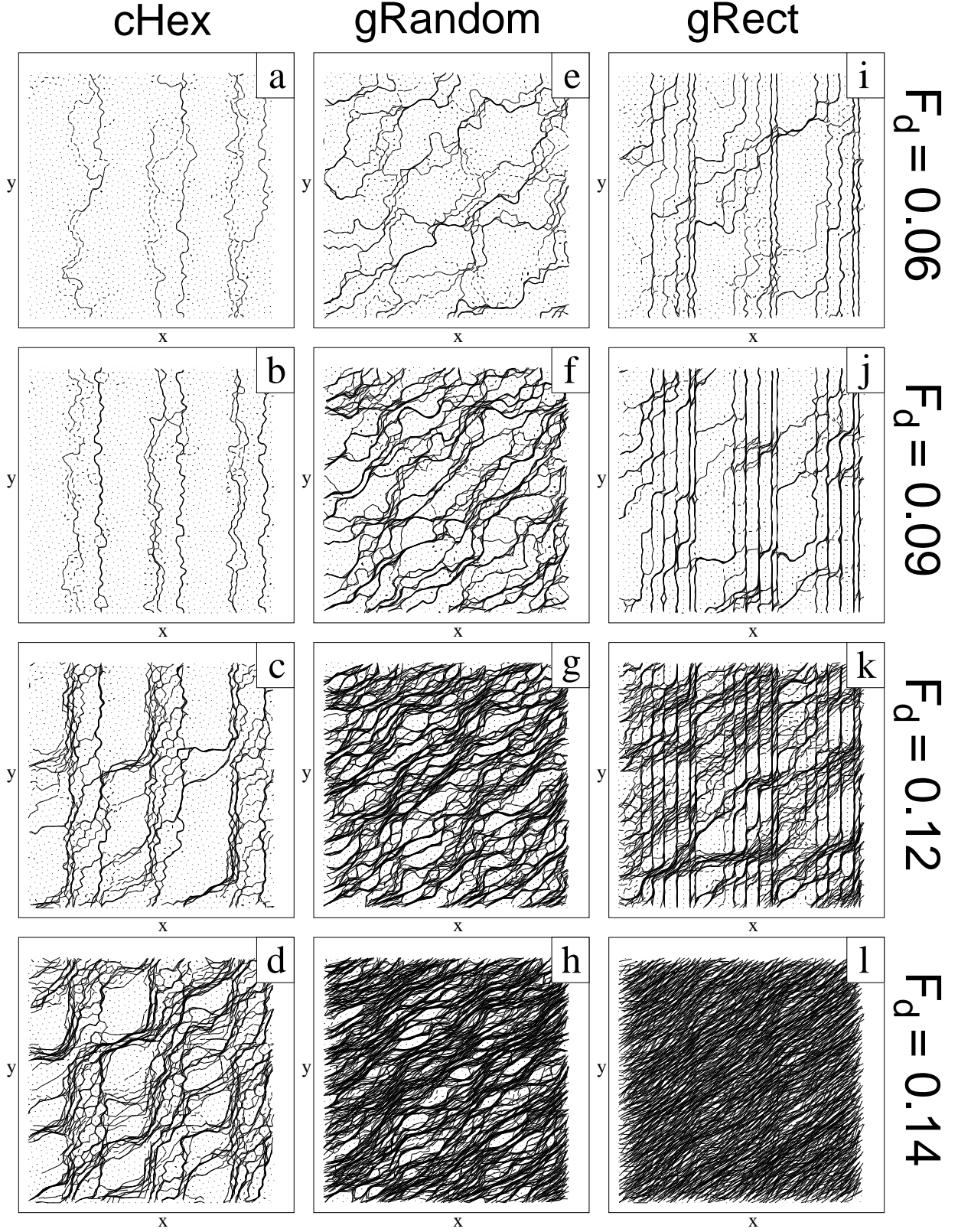}
\caption{
Lines: vortex trajectories, dark dots: vortex positions, and 
light dots: pin positions for
selected values of $F_d$
in the current-driven systems described in Fig.~\ref{fig:17}
with $F_p=0.55$, $H/H_\phi=1.4$, and drive orientation 
$\theta=30^{\circ}$.
(a,b,c,d): cHex; (e,f,g,h): gRandom;  (i,j,k,l): gRect. 
$F_d=$ (a,e,i) 0.06, (b,f,j) 0.09, (c,g,k) 0.12,
and (d,h,l) 0.14.
}
\label{fig:18}
\end{figure*}

Next we consider the effect of changing $\theta$, the orientation of the
driving direction,
for various gradient pinning arrays. 
In Fig.~\ref{fig:17} we plot the average velocity 
in the $x$-direction $\langle V_{x} \rangle$ and $y$-direction 
$\langle V_{y} \rangle$ as a function of $F_d$
for the cHex, gRandom, and gRect arrays, 
at a drive angle of $\theta = 30^{\circ}$. 
For the cHex array in Fig.~\ref{fig:17}(a), 
$\langle V_{y} \rangle$ is nonzero for $0.05 < F_{d} < 0.10$ 
while $\langle V_{x} \rangle$ remains zero, 
meaning that vortex flow occurs strictly in the $y$-direction,
perpendicular to the gradient, 
even though the largest component of the external drive is along
the $x$ direction.
Once the vortices start to move along the $x$ direction as well,
we find a crossing of the $\langle V_x\rangle$ and $\langle V_y\rangle$
curves near $F_d=0.15$.
In Fig.~\ref{fig:18}(a,b,c,d) we illustrate the vortex flow in the cHex
array at
the drive values marked with the letters {\it a-d} in Fig.~\ref{fig:17}(a).
At $F_{d}=0.06$ and $0.09$ in Figs.~\ref{fig:18}(a) and (b), 
the vortex flow occurs in winding rivers aligned with the $y$-direction, 
which pass through the portions of the sample in which the pinning density
is lowest.
The cHex array 
exhibits a pronounced guidance or channeling effect since 
most of the pins are occupied in the higher pinning density regions, 
creating a barrier to vortex motion,  
while the local depinning threshold is reduced in areas of lower
pinning density. 
At $F_{d} = 0.12$, Fig.~\ref{fig:18}(c) shows that vortices begin
to cross between the easy flow channels, passing through the
denser pinning region.
This corresponds with the emergence of a
non-zero value of $\langle V_{x} \rangle$ 
in Fig.~\ref{fig:17}(a) at point c.
At higher drives the amount of motion in the $x$-direction 
between easy-flow channels 
increases, as illustrated in Fig.~\ref{fig:18}(d) at $F_{d} = 0.14$.

The depinning transition falls at a much lower value of $F_d$
for the random gradient array, as shown 
in Fig.~\ref{fig:17}(b),
and unlike in the cHex array there is no vortex guidance effect in
which flow
only occurs in the $y$-direction.
Instead, $\langle V_x\rangle$ and $\langle V_y\rangle$ both
increase smoothly with increasing $F_d$, and
there is no crossing of the velocity curves. 
In Fig.~\ref{fig:16} we showed 
that a random gradient array driven in the 
gradient direction with $\theta=0^{\circ}$ 
has essentially identical transport properties 
to a uniform random array, indicating that the addition of a gradient to
a random array does not substantially affect vortex flow. 
Similarly, we find that for $\theta \neq 0^{\circ}$ the flow 
through random arrays is also not affected by the addition of a gradient.
In Fig.~\ref{fig:18}(e,f,g,h) we show the vortex trajectories for
the drives labeled {\it e-h} in Fig.~\ref{fig:17}.
At $F_{d} = 0.06$ and $0.09$ in Fig.~\ref{fig:18}(e) and (f), the vortices
move in both the $x$ and $y$ directions and form winding channels
oriented along the driving direction.
Due to the less effective pinning in the gRandom array, 
the number of moving vortices 
is much greater 
than in the
cHex array at the same drive, as can be seen by comparing
the high trajectory
densities in Fig.~\ref{fig:18}(e,f) to the 
much sparser trajectories in Fig.~\ref{fig:18}(a,b).
At $F_{d} = 0.14$ in Fig.~\ref{fig:18}(h),
most of the vortices are moving.
At drives higher than those illustrated here, 
the vortices dynamically 
reorder in a manner similar to the
dynamic reordering transitions observed for vortices driven
over uniform random pinning arrays \cite{54,55}.

Figure \ref{fig:17}(c) shows $\langle V_x\rangle$ and $\langle V_y\rangle$
versus $F_d$ curves for the
rectangular gradient array. 
Here there is a pronounced guidance effect for $0.02 < F_{d} < 0.11$; 
however, unlike in the cHex array, $\langle V_{x} \rangle$ is not strictly
zero, indicating that some vortex motion 
is occurring in the $x$-direction between the 1D columns of pins
since the gRect array is less effective in restricting vortex motion
along its gradient. 
The vortex flow images in Fig.~\ref{fig:18}(i,j,k,l) 
correspond to the labeled drives in Fig.\ref{fig:17}(c). 
For $F_{d} = 0.06$ and $0.09$ in Fig.~\ref{fig:18}(i) and (j), 
the vortices follow nearly 1D paths along the $y$ direction with
very little winding, while there are a smaller number of
flow channels
oriented approximately parallel to the $x$ direction that 
connect some of the 1D paths.
The 1D vertical flow channels are concentrated in the  
least dense portions of the pinning array, 
and are similar in nature to the 1D flow of interstitial vortices 
observed in uniform square or rectangular periodic pinning arrays \cite{56}.  
At $F_{d} = 0.12$ in Fig.~\ref{fig:18}(k),
there is a significant increase in the amount of
vortex flow along the $x$-direction,
while for $F_{d} = 0.14$ in Fig.~\ref{fig:18}(l)
the flow is continuous though the system and oriented
with the driving direction.
As with the gRandom array, the high trajectory densities indicate 
pinning which is less effective than in the cHex array. 
For $F_{d} > 0.20$ the vortices dynamically order, as indicated by
the onset of a linear regime of $\langle V_x\rangle$ versus $F_d$
in Fig.~\ref{fig:17}(c).

\begin{figure}
\includegraphics[width=\columnwidth]{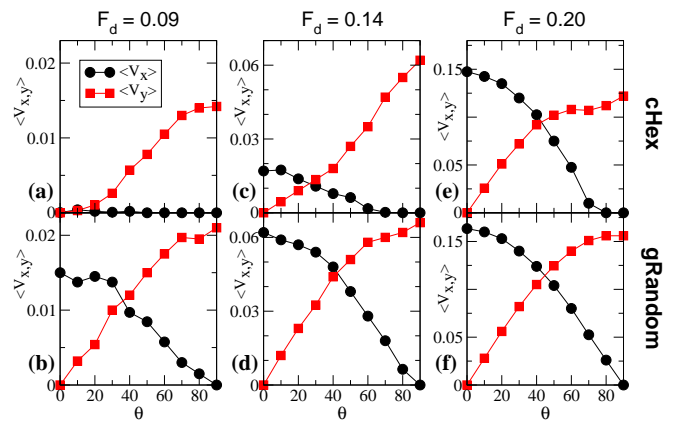}
\caption{
$\langle V_{x} \rangle$ (circles) 
and $\langle V_{y} \rangle$ (squares) 
vs drive angle $\theta$ 
for the cHex (a,c,e) and gRandom (b,d,f) arrays,  
at $F_{d} = 0.09$ (a,b), $F_{d} = 0.14$ (c,d), and $F_{d} = 0.20$ (e,f). 
We fix $F_{p} = 0.55$ and $H/H_{\phi} = 1.4$.   
}
\label{fig:19}
\end{figure}

We summarize the transport properties for different drive angles 
in Fig.~\ref{fig:19}, 
where we plot $\langle V_{x} \rangle$ 
and $\langle V_{y} \rangle$ versus $\theta$ for different
values of $F_d$ in the cHex and gRandom arrays.
In the absence of pinning, we would obtain
$\langle V_{x} \rangle = F_{d}\cos(\theta)$ and 
$\langle V_{y} \rangle = F_{d}\sin(\theta)$. 
For the cHex array with $F_{d} = 0.09$, shown in Fig.~\ref{fig:19}(a), 
$\langle V_{x} \rangle \approx 0$ for all $\theta$, 
while $\langle V_{y} \rangle$ 
monotonically increases with increasing $\theta$.  
This drive is too weak to depin the vortices along the $x$-direction,  
but the vortices can flow in the $y$-direction through the low density
regions as was shown in Fig.~\ref{fig:18}(b). 
Fig.~\ref{fig:19}(b) shows the same quantities for the gRandom
array at the same drive $F_d=0.09$.
Here $\langle V_{x} \rangle$ is nonzero over the whole range of $\theta$
since the depinning threshold in the $x$-direction is much lower 
for the gRandom array.
There is an asymmetry in the $x$ and $y$ direction responses,
as indicated by the fact that 
$\langle V_{x} \rangle = 0.015$ at $\theta = 0^{\circ}$ 
but $\langle V_{y} \rangle = 0.021$ at $\theta = 90^{\circ}$.
This indicates that  
the pinning in the moving phase is slightly more effective 
when the vortices are moving along the gradient direction.
However, the vortex velocities are much larger in the gRandom array
than in the cHex array over the whole range of $\theta$,
showing the robust enhanced effectiveness of the conformal pinning. 

When the drive in the cHex array is increased to $F_{d} = 0.14$, 
Fig.~\ref{fig:19}(c) shows that the $x$-direction depinning threshold 
is exceeded for $\theta < 70^{\circ}$, but 
that the vortex motion along the $x$ direction is still greatly 
suppressed even when the vortices are flowing.
There is still a strong vortex channeling effect 
in the $y$-direction, as reflected in the large asymmetry 
between $\langle V_{x} \rangle (\theta)$ 
and $\langle V_{y} \rangle (90^{\circ}-\theta)$. 
For the gRandom array at $F_d=0.14$, Fig.~\ref{fig:19}(d) indicates
that this drive magnitude is already large enough to render the 
pinning mostly ineffective, as also shown in Fig.~\ref{fig:18}(h). 
The $\langle V_{x} \rangle$ and $\langle V_{y} \rangle$ curves 
are almost mirror-symmetric, indicating that the pinning gradient has
little effect on the motion of the vortices.
At $F_{d} = 0.20$, 
Fig.~\ref{fig:19}(e) shows that the response of cHex becomes increasingly
symmetrical, although there is still a small channeling 
regime for $\theta \geq 80^{\circ}$ where the
flow is strictly confined along the $y$-direction. 
At higher drives (not shown), the $\langle V_{x} \rangle$ and 
$\langle V_{y} \rangle$ curves for cHex become symmetric. 
For the gRandom array at $F_d=0.20$,
Fig.~\ref{fig:19}(f) indicates that the response is almost
completely symmetrical.

\section{Summary}

We have investigated the magnetization, ordering, and transport for
vortices interacting with pinning arrays where there is a gradient in
the pinning density. 
We consider
conformal
pinning arrays constructed by taking a conformal transformation of a
uniform pinning array; the new structure has a density gradient
but preserves certain crystalline topological features of the original
lattice. In particular, we examine conformal crystals obtained from
uniform hexagonal, square, Archimedean and quasiperiodic arrays.
We also investigate selected non-conformal arrays with a pinning gradient, 
including a rectangular pinning array with a 1D gradient and a
random arrangement of pinning sites with a gradient. 
In general we find that conformal arrays 
which are transformed versions of 
uniform pinning with well-defined commensurability peaks 
give the strongest pinning over the widest range of fields. 
The hexagonal and square conformal arrays produce the strongest pinning
and highest critical currents,
followed by the conformal Archimedean and conformal Penrose arrays. 
The random gradient array gives only a slight enhancement in pinning over
a uniform random array, indicating that it is the
preservation of the local periodicity in the conformal transformation
combined with the gradient that gives rise to the enhanced pinning.
We show that for the conformal crystal arrays, a portion
of the vortices in the sample is locally commensurate 
with a portion of the pinning array, and this local commensuration
effect gradually moves through the system as the externally
applied magnetic field is changed. 
When the field becomes high enough 
that the vortex density exceeds the pinning density at the
outer edge of the sample, the effectiveness of the conformal arrays 
breaks down, although they
still show a small enhancement in pinning compared to random arrays. 
The rectangular gradient pinning array 
has relatively weak pinning compared to the
conformal arrays at lower fields; however, for higher fields, 
it develops a pronounced broad peak due
to a novel commensuration effect 
in which integer numbers of ordered columns of
vortices fit between the columns of pinning sites.

We also show that the enhanced pinning by the conformal arrays can 
be observed in current-driven transport measurements. 
Compared to uniform random or random gradient arrays, 
the conformal arrays have higher critical depinning forces, 
and lower average vortex velocities in the moving phase.
When the vortices are driven at different
angles with respect to the gradient direction, 
the conformal arrays guide vortex flow perpendicular to the gradient, 
in channels that pass through the regions of lower pinning density.  
The rectangular gradient array exhibits a similar guidance effect, 
with a strongly 1D flow of vortices in the interstitial regions. 

In this work we focus on vortices in gradient arrays; 
however, we expect similar
results for other systems featuring particles moving over substrates 
where the substrate has some type of gradient.

\acknowledgments
We thank Boldizs\'ar Jank\'o for useful discussions. 
This work was carried out under the auspices of the 
NNSA of the 
U.S. DoE
at 
LANL
under Contract No.
DE-AC52-06NA25396.

\end{document}